\documentclass[11pt,a4paper]{article}
\usepackage{jheppub}
\usepackage[T1]{fontenc}
\usepackage{amssymb}
\usepackage{mathtools}
\usepackage{stackengine}
\usepackage{scalerel}

\usepackage{hyperref}
\usepackage{epsfig}
\usepackage{amsmath,latexsym,amssymb}
\usepackage{graphicx}
\usepackage{braket}
\usepackage{pdfsync}

\usepackage{framed}
\usepackage{mathtools}

\usepackage{jheppub}
\usepackage[T1]{fontenc}
\usepackage{amssymb}
\usepackage{mathtools}
\usepackage{amsmath}
\usepackage{bm}
\usepackage{stackengine}

\usepackage{scalerel}
\usepackage{eufrak}
\usepackage{framed}

\usepackage{mathtools}

\usepackage{pstricks}

\usepackage{amsmath}
\usepackage{bbold}
\usepackage{bm}
\usepackage{latexsym}
\usepackage{braket}
\usepackage{slashed}
\usepackage{graphicx,booktabs,multirow}
\usepackage{eufrak}
\numberwithin{equation}{section}

\usepackage{color}
\usepackage{pstricks}
\usepackage{amssymb}

\makeatletter
\newcommand{\doublewidetilde}[1]{{%
  \mathpalette\double@widetilde{#1}%
}}
\newcommand{\double@widetilde}[2]{%
  \sbox\z@{$\m@th#1\widetilde{#2}$}%
  \ht\z@=.9\ht\z@
  \widetilde{\box\z@}%
}
\makeatother

\usepackage{hyperref}
\usepackage{slashed}
\usepackage{epsfig}
\usepackage{amsmath,latexsym,amssymb}
\usepackage{graphicx}
\usepackage[latin1]{inputenc}
\usepackage{braket}
\usepackage{pdfsync}

\usepackage{framed}

\usepackage{mathtools}

\def\be{\begin{equation}}
\def\ee{\end{equation}}
\def\ba{\begin{eqnarray}}
\def\ea{\end{eqnarray}}

\newcommand{\bz}{\bar{z}}

\newcommand{\bh}{\bar{h}}

\newcommand{\comment}[1]{}

\newcommand{\eea}{\end{eqnarray}}

\author{
Stephan Stieberger${}^{1}$, Tomasz R.\ Taylor${}^{2,3}$,\, Bin Zhu${}^{4}$\\[0.5cm]
$^1${\it Max--Planck--Institut f\"{u}r Physik,	Werner--Heisenberg--Institut, 80805 M\"unchen, Germany}\\
 $^2${\it Department of Physics,
  Northeastern University, Boston, MA 02115, USA}\\
  $^3${\it Faculty of Physics, University of Warsaw, ul. Pasteura 5, 02-093 Warsaw, Poland}\\
$^4${\it School of Mathematics and Maxwell Institute for Mathematical Sciences,\\ University of Edinburgh,
EH9 3FD, UK }\\[0.2cm]
}

\emailAdd{stephan.stieberger@mpp.mpg.de}
\emailAdd{taylor@neu.edu}
\emailAdd{bzhu@exseed.ed.ac.uk}

\title{Yang-Mills as a Liouville Theory}

\abstract{We propose a description of the gluon scattering amplitudes as the inverse Mellin transforms of the conformal correlators of light operators in two-dimensional Liouville theory tensored with WZW-like chiral currents on the celestial sphere. The dimensions of operators are Mellin dual to gluon light cone energies while  their positions are determined by the gluon momentum directions. Tree-level approximation in Yang-Mills theory corresponds to the semiclassical limit of Liouville theory. By comparing  subleading corrections, we find $b^2= (8\pi^2)^{-1}\beta_0 \,g^2(M)$, where $b$ is the Liouville coupling constant, $g(M)$ is the Yang Mills coupling at the renormalization scale $M$ and $\beta_0$ is the one-loop coefficient of the Yang-Mills beta function.}

\notoc
\makeatletter
\gdef\@fpheader{}
\makeatother

\begin{document}
\maketitle
\noindent \section{}\vskip -1cm
In a recent paper  \cite{Stieberger:2022zyk}, we established an intriguing connection between the tree-level gluon scattering amplitudes and the correlators of two-dimensional Liouville theory on the celestial sphere. The gluon amplitudes were evaluated in the presence of a dilaton source and transformed into ``celestial''  amplitudes \cite{Pasterski:2017kqt,Pasterski:2017ylz} by taking Mellin transforms with respect to the light cone energies of scattered gluons. The dimensions of Liouville operators were Mellin duals of such energies.
Their positions were determined by the celestial map between the directions of light-like momenta and points on two-dimensional celestial sphere.\footnote{See reviews of celestial holography in Refs.\cite{Strominger:2017zoo,Raclariu:2021zjz,Pasterski:2021rjz,Pasterski:2021raf}. Most of the recent work has focused on extracting CFT data of the putative celestial CFT from scattering amplitudes in four dimensions, e.g., celestial OPEs \cite{Fan:2019emx,Pate:2019lpp}, infinite-dimensional algebras \cite{Fotopoulos:2019vac,Guevara:2021abz,Strominger:2021lvk}, differential equations \cite{Banerjee:2020zlg, Banerjee:2020vnt, Hu:2021lrx}, and connections to twistor theory \cite{Adamo:2019ipt,Adamo:2021lrv,Adamo:2021zpw, Costello:2022wso,Bittleston:2023bzp}.}
The celestial amplitudes matched the Liouville correlators evaluated in the limit of small Liouville coupling, $b\to 0$, which corresponds to the infinite central charge limit.
This construction has been recently generalized in Ref.\cite{Taylor:2023bzj}
 to celestial amplitudes in $\mathcal{N}=1$ supersymmetric Yang-Mills theory coupled to dilatons.

In the present work we proceed in the opposite direction. We start from the operators associated with gluons, constructed as the products of holomorphic Wess-Zumino-Witten (WZW) curents times the so-called light Liouville operators. The current part carries the information about gluon gauge charges and spins. The Liouville part determines their dimensions. The three-point correlation functions of such operators factorize into a relatively simple, exactly known WZW correlators times the three-point correlators of light Liouville operators. The latter ones are known exactly from DOZZ formula \cite{Dorn:1994xn,Zamolodchikov:1995aa} and can be expressed in terms of Zamolodchikovs' $\Upsilon$ function \cite{Zamolodchikov:1995aa}. We perform inverse Mellin transformations on the two-dimensional correlators. By using the celestial map, we construct the corresponding gluon scattering amplitudes.
We can recover the gluon amplitudes, at the tree level and beyond, without the dilaton background, by taking the limit of inverse Mellin transforms in which the dilatons decouple. This procedure can be performed exactly at the leading order in the Liouville coupling ($b\to 0$), corresponding to the tree level approximation in Yang-Mills theory. We also go beyond the leading order and identifiy
some corrections pointing towards a direct relation between the Liouville and Yang-Mills couplings.

 The Lagrangian density of two-dimensional Liouville theory is given by
\be{\cal L}= \frac{1}{\pi}{\partial \phi\over\partial z}{\partial\phi\over\partial\bz}+\mu e^{2b\phi}\ ,\ee
where $z$ and $\bz$ are the complexified (Euclidean) spacetime coordinates, $b$ is the dimensionless Liouville coupling constant and $\mu$ is the ``cosmological constant'' scale parameter.
The theory has a ``background charge at infinity,''
\be Q=b+\frac{1}{b},\ee
which is related to the central charge by
\be c=1+6Q^2.\label{cch}\ee
The ``light'' primary field operators have the form:
\be V_\sigma(z,\bz)=e^{2\sigma b \phi(z,\bz)},\label{lops}\ee
with the exponents parametrized by $b$-independent parameters $\sigma$. Their conformal dimensions are given by
\be d(\sigma)=2\sigma+2b^2\sigma(1-\sigma)\ .\label{dims}\ee

We introduce spin and gauge charges into the two-dimensional system by including a WZW-like holomorphic sector. The WZW current $J^a(z)$, with $a$ labeling the adjoint representation of the Lie group, has chiral weights
$(h,\bh)=(1,0)$. We also include another operator in the adjoint representation, $\widehat J^a(z)$, with $(h,\bh)=(-1,0)$. The only property of this chiral system\footnote{For a more  detailed discussions of this chiral system, see Ref.\cite{Costello:2022wso}.}  relevant to our discussion is the form of the three-point correlator
\be \big\langle \widehat J^{a_1}(z_1)\widehat J^{a_2}(z_2) J^{a_3}(z_3)\big\rangle=f^{a_1a_2a_3}\frac{z_{12}^3}
{z_{23}z_{31}},\ee
where $z_{ij}=z_i-z_j$ and $f^{a_1a_2a_3}$ are the structure constants.

We construct the operators associated with the positive helicity gluons in the following way:
\be
O^{+a}_{\Delta}(z,\bar{z}) =  F_+(\Delta,\mu, b)\,
J^a(z) e^{2\sigma(\Delta-1)b\phi(z,\bar{z})} \, ,
\ee where $F_+(\Delta,\mu, b)$ is a normalization factor
and $2\sigma(\Delta-1)$ ensures dimension $\Delta-1$ of the Liouville operator. At the leading order ${\cal O}(b^0)$, $2\sigma(\Delta-1)=\Delta-1$.
Similarly, for the negative helicity gluon,
\be
O^{-a}_{\Delta}(z,\bar{z}) =  F_-(\Delta,\mu, b)\,
J^a(z) e^{2\sigma(\Delta+1)b\phi(z,\bar{z})} \, ,
\ee
Note that the normalization factors $F_{\pm}(\Delta,\mu, b)$ depend on the dimensions $\Delta$, therefore they contribute to inverse Mellin transforms in a nontrivial way.

We are interested in the ``MHV'' correlator
\ba\
&\hskip-9ex \big\langle O^{-a_1}_{\Delta_1}(z_1,\bar{z}_1)
O^{-a_2}_{\Delta_2}(z_2,\bar{z}_2)O^{+a_3}_{\Delta_3}(z_3,\bar{z}_3)
\big\rangle=f^{a_1a_2a_3}\displaystyle\frac{z_{12}^3}
{z_{23}z_{31}}F_{1-}F_{2-}F_{3+}\!\times~~~~~~~~\\[2mm] &\times (z_{12} \bar{z}_{12})^{\frac{\Delta_3-\Delta_1-\Delta_2-3}{2}} (z_{23}\bar{z}_{23})^{\frac{\Delta_1-\Delta_2-\Delta_3+1}{2}} (z_{13}\bar{z}_{13})^{\frac{\Delta_2-\Delta_1-\Delta_3+1}{2}}\times
C(\alpha_1,\alpha_2,\alpha_3)\ ,\nonumber
\ea
where the three-point Liouville coefficient is given by the famous DOZZ formula \cite{Dorn:1994xn,Zamolodchikov:1995aa}:
\begin{align}
C(\alpha_1,\alpha_2, \alpha_3) =& \Big[ \pi \mu \gamma(b^2) b^{2-2b^2}\Big]^{(Q-\sum \alpha_i)/b}
\times\\
& \frac{\Upsilon_0 \Upsilon(2\alpha_1) \Upsilon(2\alpha_2)\Upsilon(2\alpha_3)}{\Upsilon(\alpha_1+\alpha_2+\alpha_3-Q)\Upsilon(\alpha_1+\alpha_2-\alpha_3)\Upsilon(\alpha_2+\alpha_3-\alpha_1)\Upsilon(\alpha_3+\alpha_1-\alpha_2)} \, ,\nonumber
\end{align} in our case
specified to the case of light operators with $\alpha_i=\sigma_ib$.
Here, $\Upsilon$ is the function defined in Zamolodchikovs' Ref.\cite{Zamolodchikov:1995aa}.

The semiclassical ($b\to 0$) limit of the three-point correlator of light Liouville fields has been studied before by Harlow, Maltz and Witten \cite{Harlow:2011ny}.
We use the following formulas from Ref.\cite{Harlow:2011ny}:
\begin{align}
\Upsilon(x-1/b) &= \gamma(x/b-1/b^2)^{-1} b^{1+2/b^2 -2x/b} \, \Upsilon(x) \, , \\
\Upsilon_0 &= \frac{\cal C}{b^{1/2}} \exp\left(-\frac{1}{4b^2} \log b + \dots \right) \, ,\\
\Upsilon_b(\sigma b) &= \frac{{\cal C} b^{1/2-\sigma}}{\Gamma(\sigma)} \exp\left( -\frac{1}{4b^2} \log b + \dots\right) \, ,
\end{align}
where $\cal C$ is a constant and $\gamma(x)=\Gamma(x)/\Gamma(1-x)$.
In this way, we find
\begin{align}
C(\sigma_1 b, \sigma_2 b, \sigma_3 b) =&  \frac{\pi \tilde{\mu} \gamma(1/b^2) \, \gamma(\sum \sigma_i-1-1/b^2)\, [\pi \mu \gamma(b^2) b^{-2b^2}]^{1-\sum \sigma_i}}{ b^5 } \\
&\times \frac{\Gamma(\sigma_1+\sigma_2+\sigma_3-1)\Gamma(\sigma_1+\sigma_2-\sigma_3)\Gamma(\sigma_2+\sigma_3-\sigma_1)\Gamma(\sigma_3+\sigma_1-\sigma_2)}{\Gamma(2\sigma_1)\Gamma(2\sigma_2)\Gamma(2\sigma_3)} \, ,\nonumber
\end{align}
where the ``dual'' cosmological constant $\tilde{\mu}$ is related to $\mu$ as follows
\begin{align}
\pi \tilde{\mu} \gamma(1/b^2) = (\pi \mu \gamma(b^2))^{1/b^2} \, .  \label{eq:mut}
\end{align}

Our goal is to apply the celestial map to the inverse Mellin  transform,
\ba
{\cal A}_{3G}(\omega_i,z_i,\bz_i)=\displaystyle
 M^{\Delta_1+\Delta_2+\Delta_3-3} \left(\frac{1}{2\pi i}\right)^3&\displaystyle\!\!\int_{c-i\infty}^{c+i\infty} d\Delta_1 d\Delta_2 d\Delta_3 \,  \omega_1^{-\Delta_1}\, \omega_2^{-\Delta_2} \, \omega_3^{-\Delta_3}~~~~~ \\[2mm] &\;\;\times\, \big\langle O^{-a_1}_{\Delta_1}(z_1,\bar{z}_1)
O^{-a_2}_{\Delta_2}(z_2,\bar{z}_2)O^{+a_3}_{\Delta_3}(z_3,\bar{z}_3)
\big\rangle\ ,\nonumber\ea
where the integrations are performed on the complex plane along the lines of real constant $c>0$; at the end, we will take the limit of $c\to 0^+$. Note that connecting two to four dimensions necessitates introducing
a ``renormalization'' scale $M$ in order to ensure the correct mass dimension $-3$ of the three-gluon amplitude.
As mentioned before, the integrands depend on the normalization constants $F_{\pm}$. We will see below that the following choice leads to the desired result in the semiclassical limit:
\begin{align}
F_+(\Delta, \mu, b)=& [\pi \mu \gamma(b^2)b^{-2b^2}]^{\sigma(\Delta-1)}
\;\Gamma[2\sigma(\Delta-1)]\ ,\\[2mm]
F_-(\Delta, \mu, b)=& [\pi \mu
\gamma(b^2)b^{-2b^2}]^{\sigma(\Delta+1)-1/2}
\;\Gamma[2\sigma(\Delta+1)]\ .
\end{align}
Then as $b\to 0$, when $2\sigma=\Delta-1$ for positive helicity gluon and $2\sigma=\Delta+1$ for negative helicity gluon, the leading term becomes
\be {\cal A}_{3G}^{(0)}(\omega_i,z_i,\bz_i)=
\frac{\pi\, \tilde{\mu}}{b\, M^2}
f^{a_1a_2a_3}\displaystyle\frac{z_{12}^3}
{z_{23}z_{31}}I^{(0)}(\omega_1,\omega_2,\omega_3)\ ,\label{azero}\ee
where
\ba
I^{(0)}(\omega_1,\omega_2,\omega_3)=&\displaystyle\!\!\!
 \hskip -9ex ~~~~~\left(\frac{1}{2\pi i}\right)^3 \,\int_{c-i\infty}^{c+i\infty} d\Delta_1 d\Delta_2 d\Delta_3 \,  M^{\Delta_1+\Delta_2+\Delta_3-1} \omega_1^{-\Delta_1}\, \omega_2^{-\Delta_2} \, \omega_3^{-\Delta_3}~~~~~~\\[1.4mm]
& \times\, \Gamma\left(\frac{\Delta_1+\Delta_2+\Delta_3-1}{2} \right) \Gamma\left(\frac{\Delta_1+\Delta_3-\Delta_2-1}{2}\right) \Gamma\left(\frac{\Delta_2+\Delta_3-\Delta_1-1}{2}\right) \Gamma\left(\frac{\Delta_1+\Delta_2-\Delta_3+3}{2}  \right) \nonumber\\ [2mm]
  &\times\,
 (z_{12} \bar{z}_{12})^{\frac{\Delta_3-\Delta_1-\Delta_2-3}{2}} (z_{23}\bar{z}_{23})^{\frac{\Delta_1-\Delta_2-\Delta_3+1}{2}} (z_{13}\bar{z}_{13})^{\frac{\Delta_2-\Delta_1-\Delta_3+1}{2}} \nonumber
\ea
It is convenient to use the integral representation
\be
\Gamma(z) = \int_0^{+\infty} dt\, e^{-t} \,  t^{z-1} \, .
\ee
to rewrite the inverse Mellin transform as
\begin{align}
I^{(0)}(\omega_1,\omega_2,\omega_3)
=&\frac{1}{M} \, \left(\frac{1}{2\pi i}\right)^3 \, \int_{c-i\infty}^{c+i\infty} d\Delta_1 d\Delta_2 d\Delta_3 \, \int_0^{+\infty} dt_0 dt_1 dt_2 dt_3 \,
e^{\Delta_1x_1}e^{\Delta_2x_2}e^{\Delta_3x_3}\label{inte0}\\
&\times \, e^{-t_0-t_1-t_2-t_3}\,\frac{t_0^{-\frac{1}{2}}\, t_1^{-\frac{1}{2}} \, t_2^{-\frac{1}{2}}\, t_3^{\frac{3}{2}}}{t_0 \, t_1\, t_2 \, t_3} (z_{12} \bar{z}_{12})^{-\frac{3}{2}} (z_{23}\bar{z}_{23})^{\frac{1}{2}} (z_{13}\bar{z}_{13})^{\frac{1}{2}}\nonumber
\end{align}
where
\begin{align}
x_1 &= \frac{1}{2}\ln\left( \frac{M^2 t_0 \, t_1 \, t_3 \, z_{23}  \bar{z}_{23}}{ \omega_1^2 \,  t_2 \, z_{12}\bar{z}_{12} \, z_{13}\bar{z}_{13}} \right)\\ x_2&= \frac{1}{2} \ln\left(\frac{M^2t_0 \, t_2 \, t_3 \, z_{13}  \bar{z}_{13}}{\omega_2^2 \, t_1 \, z_{12}\bar{z}_{12} \, z_{23}\bar{z}_{23}} \right)\\
x_3 &= \frac{1}{2} \ln \left( \frac{M^2 t_0 \, t_1 \, t_2 \, z_{12}  \bar{z}_{12}}{\omega_3^2 \, t_3 \, z_{23}\bar{z}_{23} \, z_{13}\bar{z}_{13}} \right) \, .
\end{align}
In terms of these variables,
\begin{align}
t_1 = \frac{ \omega_1\, \omega_3 \,e^{x_1+x_3} \, |z_{13}|^2}{M^2 t_0}\ ,~  t_2 = \frac{ \omega_2 \, \omega_3 \,e^{x_2+x_3}\,  |z_{23}|^2}{M^2 t_0}\ ,~
t_3 = \frac{ \omega_1 \, \omega_2 \,e^{x_1+x_2} \, |z_{12}|^2}{ M^2t_0}\ .
\end{align}
After changing the integration variables from $t_1,t_2,t_3$ to $x_1,x_2,x_3$ and performing inverse Mellin transforms, we obtain
\be I ^{(0)}(\omega_1,\omega_2,\omega_3)=
\frac{2\,\omega_1\omega_2}{\omega_3M^2}  \int_0^{+\infty} \, dt_0 \, e^{-t_0 -\frac{Q^2}{M^2 t_0} }\, t_0^{-2}\ .\label{izero}
\ee
where
\be
Q^2 =  \omega_1 \omega_2 |z_{12}|^2 +\omega_1 \omega_3 |z_{13}|^2 +\omega_2 \omega_3 |z_{23}|^2 \, .
\ee

According to the celestial map,
\be \sqrt{\omega_i\omega_j}z_{ij}=\langle i j\rangle\ ,\qquad \omega_i \omega_j |z_{ij}|^2=2p_i\cdot p_j\ ,\ee
therefore
\be
Q^2 =(p_1+p_2+p_3)^2\ ,\ee
and $Q=p_1+p_2+p_3$ can be identified as the total momentum of the gluon system.
After inserting the result (\ref{izero}) into Eq.(\ref{azero}) and using
\be
\int_0^{+\infty}  \, dt_0 \, e^{-t_0 -\frac{ Q^2}{M^2 t_0} }\, t_0^{-2} = 2\sqrt{\frac{M^2}{Q^2}} \,K_1\left(2
\sqrt{\frac{Q^2}{M^2}}\right) \, ,\label{besin}
\ee
where $K_1$ is a modified Bessel function,
we obtain
\be {\cal A}_{3G}^{(0)}(\omega_i,z_i,\bz_i)=
\frac{4\pi\, \tilde{\mu}}{b\, M^4}
f^{a_1a_2a_3}\displaystyle\frac{\langle 12\rangle^3}
{\langle 23\rangle\langle 31\rangle}\sqrt{\frac{M^2}{Q^2}} \, K_1\left(2
\sqrt{\frac{Q^2}{M^2}}\right). \label{afin}\ee
Note that Bessel integrals (\ref{besin}) had already appeared in AdS amplitudes \cite{Penedones:2010ue}. Here they appear in the inverse Mellin transform of the WZW-Liouville correlator (\ref{afin}), which at this point seems to be different from the three-gluon amplitude of Ref.\cite{Stieberger:2022zyk} evaluated in Minkowski space. In the latter case, the amplitude  was evaluated in the presence of a dilaton background, which was taken into account by one insertion of the dilaton source. It contained the pole $(Q^2)^{-1}$ originating from the massless dilaton propagator connecting the source to the gluon system. The single source approximations, however, can be justified only in the limit of small $Q^2$. In this limit,
the Bessel function can be expanded as
\begin{align}
2\sqrt{\frac{M^2}{Q^2}} \, K_1\left(2\sqrt{\frac{Q^2}{M^2}}\right) = \frac{M^2}{ Q^2} +\dots \, ,
\end{align}
therefore
\be {\cal A}_{3G}^{(0)}(\omega_i,z_i,\bz_i)=
\frac{2\pi\, \tilde{\mu}}{b M^2Q^2}
f^{a_1a_2a_3}\displaystyle\frac{\langle 12\rangle^3}
{\langle 23\rangle\langle 31\rangle}+\dots \label{afin2}\ee
We want to match this correlator with the tree-level amplitude
\be {\cal A}_{3G}^{(0')}(\omega_i,z_i,\bz_i)=
\frac{g}{\Lambda\Lambda'}
f^{a_1a_2a_3} {1\over Q^2}\displaystyle\frac{\langle 12\rangle^3}
{\langle 23\rangle\langle 31\rangle}+\dots \label{afin3}\ ,\ee
where $g$ is the Yang-Mills coupling constant, $\Lambda^{-1}$ is the canonical  coupling of the dilaton to the gauge field strength and $\Lambda'$ determines the strength of the point-like dilaton source, ${\cal J}(x)=\delta^{(4)}(x)/\Lambda'$. The semiclassical limit of the Liouville correlator is equal to the tree-level amplitude provided that the Yang-Mills and dilaton parameters are related to the Liouville parameters and the renormalization scale in the following way:
\be \frac{gM^2}{\Lambda\Lambda'}=\frac{2\pi\, \tilde{\mu}}{b }\ .\label{relat}\ee

The relation between Liouville correlators and Yang-Mills amplitudes can be extended beyond the semiclassical limit. The limit of $Q^2\to 0$ singles out gluon amplitudes with one insertion of the dilaton source. These amplitudes contain the dilaton propagator and the coupling of the off-shell dilaton to the gluon system. It is well known, however, that the dilaton decouples in the zero-momentum limit \cite{Ademollo:1975pf,Shapiro:1975cz,DiVecchia:2015jaq}. Namely, the Feynman matrix element with one zero momentum dilaton is given by the Feynman matrix element evaluated in the absence of dilatons -- in our case in pure Yang-Mills theory. This observation leads to\\[2mm]
{\it Proposition:}
\ba
{\cal M}_{3G}(\omega_i,z_i,\bz_i)=\displaystyle\lim_{Q\to 0}&\displaystyle\displaystyle \frac{Q^2}{(2\pi i)^3}
\displaystyle\!\!\int_{c-i\infty}^{c+i\infty} d\Delta_1 d\Delta_2 d\Delta_3 \,  M^{\Delta_1+\Delta_2+\Delta_3-1}\omega_1^{-\Delta_1}\, \omega_2^{-\Delta_2} \, \omega_3^{-\Delta_3}~~~~~ \nonumber\\[2mm] &\;\;\times\, \big\langle O^{-a_1}_{\Delta_1}(z_1,\bar{z}_1)
O^{-a_2}_{\Delta_2}(z_2,\bar{z}_2)O^{+a_3}_{\Delta_3}(z_3,\bar{z}_3)
\big\rangle\ ,\label{propos}\ea
where ${\cal M}_{3G}$ is the {\it exact\/} three-gluon MHV Feynman matrix element (of mass dimension 1) in Yang-Mills theory.
The equation should be supplemented with a prescription how to replace two-dimensional Liouville parameters on the r.h.s.\ by four-dimensional Yang-Mills parameters on the l.h.s. All what we can extract at the leading perturbative order is written in Eq.(\ref{relat}). We need an exact and more direct relation, however, between Liouville and Yang-Mills couplings. It can be extracted by going beyond the leading order on the Yang-Mills and Liouville sides of Eq.(\ref{propos}).

In Yang-Mills theory, next-to-leading corrections originate from one-loop diagrams and are of order ${\cal O}(g^2)$ as compared to the tree level. In Liouville theory, they are of order
 ${\cal O}(b^2)$ and originate from various sources. First of all, the $\Upsilon$ function has been expanded in Ref.\cite{Harlow:2011ny} to the order ${\cal O}(b^2\ln b^2)$ only, and more work is needed to reach higher precision. Furthermore, there is a similar uncertainity in the normalization factors $F_{\pm}$. In addition, DOZZ formula is written in terms of the exponents $\sigma_i$ while the inverse Mellin transforms involve integrations over the dimensions $\Delta_i$. Eq.(\ref{dims}) implies that at the subleading order
\begin{align}
\sigma_1 &= \frac{\Delta_1+1}{2} + \frac{b^2}{4} (\Delta_1+1)(\Delta_1-1)  \nonumber\\
\sigma_2 &= \frac{\Delta_2+1}{2} + \frac{b^2}{4} (\Delta_2+1)(\Delta_2-1) \nonumber\\
\sigma_3 &= \frac{\Delta_3-1}{2} + \frac{b^2}{4} (\Delta_3-1)(\Delta_3-3)   \label{eq:sigi}
\end{align}
We leave full analysis of subleading Liouville corrections to future work, nevertheless already at this point, we can get a preliminary insight by discussing some  consequences of Eq.(\ref{eq:sigi}).

After repeating the steps leading to Eq.(\ref{inte0}), but now with the exponents related to dimensions by Eq.(\ref{eq:sigi}), we obtain
\begin{align}
I^{(1)}(\omega_1,\omega_2,\omega_3)
=&\frac{1}{M}\,\left(\frac{1}{2\pi i}\right)^3 \,\int_{c-i\infty}^{c+i\infty} d\Delta_1 d\Delta_2 d\Delta_3 \, \int_0^{+\infty} dt_0 dt_1 dt_2 dt_3 \,
e^{\Delta_1x_1}e^{\Delta_2x_2}e^{\Delta_3x_3}\nonumber\\
&\times \, e^{-t_0-t_1-t_2-t_3}\,\frac{t_0^{-\frac{1}{2}}\, t_1^{-\frac{1}{2}} \, t_2^{-\frac{1}{2}}\, t_3^{\frac{3}{2}}}{t_0 \, t_1\, t_2 \, t_3} (z_{12} \bar{z}_{12})^{-\frac{3}{2}} (z_{23}\bar{z}_{23})^{\frac{1}{2}} (z_{13}\bar{z}_{13})^{\frac{1}{2}}
\\
&\hskip -6ex\times  \left( \frac{t_0\, t_1\, t_3}{t_2}\right)^{\frac{b^2}{4} (\Delta_1+1)(\Delta_1-1)} \left( \frac{t_0\, t_2\, t_3}{t_1}\right)^{\frac{b^2}{4} (\Delta_2+1)(\Delta_2-1)} \left( \frac{t_0\, t_1\, t_2}{t_3}\right)^{\frac{b^2}{4} (\Delta_3-1)(\Delta_3-3)}
\nonumber
\end{align}
The difference between the present case and Eq.(\ref{inte0}) is that the integrals over dimensions $\Delta_i$ become Gaussian instead of delta functions. After performing these integrals and changing the
variables from $t_1,t_2,t_3$ to $x_1,x_2,x_3$, we obtain
\begin{align}
&I^{(1)}(\omega_1,\omega_2, \omega_3)=\frac{2 \,\omega_1\omega_2}{M^2\,\omega_3} \;
 e^{\big[-\frac{b^2}{4} \ln(\frac{2\, p_1\cdot p_2}{M^2})-\frac{b^2}{4} \ln(\frac{2\, p_2\cdot p_3}{M^2})-\frac{b^2}{4} \ln(\frac{2\, p_1 \cdot p_3}{M^2})\big]}\nonumber\\
& ~~~~\times \int_{-\infty}^{+\infty} dx_1 dx_2 dx_3\, \prod_{i=1}^{3}\frac{1}{\sqrt{\epsilon_i(x_i)}}e^{\frac{-x_i^2}{\epsilon_i(x_i)}+x_i(1-\frac{b^2}{2})}\\[2mm]
&     ~~~~~\times\int_0^{+\infty}dt_0\, t_0^{-2}\,e^{-t_0 -\frac{e^{x_1+x_3} \, \omega_1\, \omega_3 \, |z_{13}|^2}{M^2 \, t_0} -\frac{e^{x_2+x_3}\, \omega_2 \, \omega_3 \,  |z_{23}|^2}{M^2 \, t_0} -\frac{e^{x_1+x_2} \, \omega_1 \, \omega_2 \, |z_{12}|^2}{M^2 t_0}} \ ,
\end{align}
where
\begin{align}
\epsilon_1(x_1)=& \pi b^2 \Big[\ln \Big(\frac{2\, p_1 \cdot p_3 \, p_1\cdot p_2}{M^2\, p_2\cdot p_3}\Big) +2 x_1\Big],\nonumber \\
\epsilon_2(x_2)=& \pi b^2 \Big[ \ln\Big (\frac{ 2\, p_2 \cdot p_3 \, p_1\cdot p_2}{M^2 \, p_1\cdot p_3}\Big) +2 x_2\Big], \\
\epsilon_3(x_3)=& \pi b^2 \Big[ \ln \Big(\frac{2\, p_1 \cdot p_3 \, p_2\cdot p_3}{M^2 \, p_1\cdot p_2}\Big) +2 x_3\Big]. \nonumber\label{eq:A3step1}
\end{align}
Since $\epsilon_{i}(x_i)\sim b^2$, we can use the expansion
\be
\frac{1}{\sqrt{4\pi \epsilon}} e^{\frac{-x^2}{4\epsilon}} = e^{\epsilon\partial_x^2} \, \delta(x) \, .
\ee
which yields the delta functions fixing $x_i=0$ at the leading order.  After expanding the remaining factors, we obtain \ba
I^{(1)}(\omega_1,\omega_2, \omega_3)=I^{(0)}(\omega_1,\omega_2, \omega_3)&\label{l1}\\[2mm]
&\displaystyle\hskip -3.2cm\times\Big[1-\frac{b^2}{4} \ln\Big(\frac{2\, p_1\cdot p_2}{M^2}\Big)-\frac{b^2}{4} \ln\Big(\frac{2\, p_2\cdot p_3}{M^2}\Big)-\frac{b^2}{4} \ln\Big(\frac{2\, p_1 \cdot p_3}{M^2}\Big)\Big]+\nonumber\dots\ .\ea
The presence of logarithmic corrections in Liouville theory indicates that the arbitrary mass scale $M$, introduced as a parameter linking Liouville and Yang-Mills theories, plays the role of renormalization scale in four dimensions. Assuming that this is indeed the case, we can extract a more precise relation between Liouville and Yang-Mills couplings by comparing Eq.(\ref{l1}) with the one-loop correction to the scattering amplitude of one dilaton with three gluons.

The one-loop corrections to the dilaton-gluon amplitudes have been studied before in
Ref.\cite{Inami:1982xt,Badger:2007si,Pereira:2022mne}.
We are interested in the ultraviolet divergent part only, which
after renormalization leads to the logarithmic running of the gauge coupling $g(Q^2)$ and of the dilaton coupling $1/\Lambda$. For three gluons \cite{Inami:1982xt,Badger:2007si,Pereira:2022mne}:
\be
\frac{g}{\Lambda}(Q^2) =  \frac{g}{\Lambda}(M^2)
 \left[ 1- \frac{3g^2(M^2)}{2(4\pi)^2}\beta_0 \, \ln \Big(\frac{Q^2}{M^2} \Big)\right] \, . \label{l2}
\ee
where $\beta_0=11c_A/3$ ($c_A$ is the Casimir operator in the adjoint representation of the gauge group) is the one-loop coefficient of the Yang-Mills beta function. By comparing the renormalization scale dependence of Eqs. (\ref{l1}) and (\ref{l2}), we find
\be b^2= \frac{\beta_0 \,g^2(M)}{8\pi^2}\ .\label{bgrel}
\ee
This relation should be taken with a grain of salt though, because it is based on a partial analysis only of the subleading Liouville corrections.

We admit that
the Proposition (\ref{propos}), together with the relation (\ref{bgrel}) contain very strong statements. Does it make sense talking about $\it exact$ gluon scattering amplitudes at all? Evidently, Yang-Mills theory confines gluons and has a mass gap. Nevertheless, gluon-like states (jets) are physically observable and according to our proposal, they are described by light Liouville operators.
Massive glueballs are probably described by some other type of operators and their amplitudes have more string-like character.

\section*{Acknowledgements}
TRT is supported by the National Science Foundation
under Grants Number PHY-1913328 and PHY-2209903, by the
NAWA Grant
``Celestial Holography of Fundamental Interactions'' and
by the Simons Foundation Grant MP-SCMPS-00001550-05.
Any opinions, findings, and conclusions or
recommendations expressed in this material are those of the authors and do not necessarily
reflect the views of the National Science Foundation. BZ is supported by the Royal Society.

\end{document}